\documentclass[12pt]{article}
\usepackage{amsmath,amssymb,color}
\usepackage{cite,epsf}
\usepackage{pstricks}
\usepackage{color}

\usepackage{graphicx,array} 
\newcommand{\be}{\begin{equation}}
\newcommand{\ee}{\end{equation}}
\newcommand{\beq}{\begin{equation}}
\newcommand{\eeq}{\end{equation}}
\newcommand{\bea}{\begin{eqnarray}}
\newcommand{\eea}{\end{eqnarray}}

\newcommand{\ba}{\begin{eqnarray}}
\newcommand{\ea}{\end{eqnarray}}

\begin{document}

\begin{titlepage}
\vspace{10pt}
\hfill
{\large\bf HU-EP-11/18}
\vspace{20mm}
\begin{center}

{\Large\bf  Wilson loop  remainder  function for null polygons \\[2mm]
in the limit of self-crossing }

\vspace{45pt}

{\large Harald Dorn, Sebastian Wuttke
{\footnote{dorn@physik.hu-berlin.de, wuttke@physik.hu-berlin.de
 }}}
\\[15mm]
{\it\ Institut f\"ur Physik der
Humboldt-Universit\"at zu Berlin,}\\
{\it Newtonstra{\ss}e 15, D-12489 Berlin, Germany}\\[4mm]

\vspace{20pt}

\end{center}
\vspace{10pt}
\vspace{40pt}

\centerline{{\bf{Abstract}}}
\vspace*{5mm}
\noindent
The remainder function of Wilson loops for null polygons becomes divergent
if two vertices approach each other. We apply RG techniques to the
limiting configuration of a contour with self-intersection. As a result
for the two loop remainder we find a quadratic divergence in the logarithm
of the distance between the two approaching vertices. The divergence is
multiplied by a factor, which is given by a pure number plus the product of two
logarithms of cross-ratios characterising the conformal geometry of
the self-crossing.
\vspace{15pt}
\noindent
\end{titlepage}
\newpage

\section{Introduction}
In recent years a lot of effort has been devoted to the investigation of 
gluon scattering amplitudes and Wilson loops in planar ${\cal N}=4$ supersymmetric
Yang-Mills theory. This includes the exposure of the  
BDS structure \cite{bds}, the relation of MHV scattering amplitudes
to Wilson loops and string surfaces at strong coupling \cite{alday-malda} and
the verification of this relation also for weak coupling, together with the analysis
of dual conformal invariance \cite{drummond}. The Wilson loops for null $n$-gons
for $ n=4,5$, via the anomalous  dual conformal Ward identity, are fixed to the
BDS structure. For $n\geq 6$ appears an additional remainder function
which depends only on conformal invariants of the corresponding polygon.
At strong coupling this remainder function can be related to the solutions of 
TBA equations for some Y-system \cite{y-syst}, but explicit analytical results
are available only for polygons in two-dimensional Minkowski space or for some
highly symmetric special cases. At weak coupling, in two loop approximation,
the remainder function for a generic hexagon has been calculated in
\cite{goncharov}  and for an octagon with restricted configurations, which can 
be embedded in two-dimensional Minkowski space, in
\cite{duhr-8}. The hexagon result has been confirmed independently 
\cite{straps} via the technique of Wilson loop operator product expansion
\cite{ope},\cite{straps}. 

In view of the complexity of the direct evaluation of the remainder functions,
useful information can come also from the study of some limiting cases
of the polygon configuration. In this sense the collinear limit
of two adjacent edges has been studied in \cite{koma} and has played a role also in
the analysis of \cite{y-syst,ope}. Another limiting case, we have in mind in this paper,
is the limit in which the polygon becomes self-crossing. As for the collinear
limit the renormalisation properties then change qualitatively. In a certain sense,
this change is even more radically, since we now have to face operator mixing
under renormalisation \cite{brandt,dorn,korch2}. 

The idea of making use of this
limit and the corresponding modified renormalisation group (RG) equation
$\grave a$ la \cite{korch1} has been developed and applied to the 
hexagon  by Georgiou in \cite{georgiou}. This paper predicted 
a singular behaviour 
$\propto \log ^3(1-u_2)$
with a pure imaginary prefactor for
the two loop hexagon remainder (depending on three cross ratios $u_1,u_2,u_3$)
in the limit $u_1=u_3,~~u_2\rightarrow 1$. It has been argued, that the origin of this term
could be related to the discontinuity of a $Li_4(1-u_2)$ term in the full remainder
function. Although there is such a $Li_4$ in the meanwhile 
available complete result of \cite{goncharov}, due to the subtleties in analytic continuation and the multi-valued nature \cite{straps} of the remainder function,
it seems to us  that still some effort is needed   to check matching of the coefficients. 
For a continuation to the Regge 
region of the corresponding scattering amplitude in the $2\rightarrow 4$
or $3\rightarrow 3$ channel see \cite{lipatov}.

Due to the null condition for the edges, for a hexagon a self-crossing can be 
realised only via crossing of two opposite edges at a common point, distinct 
from the vertices. For such crossing edges there is no characteristic free 
adjustable  conformal invariant. This explains the appearance of a pure
numeric prefactor of the $\log ^3$ divergence in \cite{georgiou}.

We expect a more interesting situation for a self-crossing of the null polygon
at a point where two vertices coincide. Then the crossing geometry exhibits
free adjustable conformal invariants. Such a situation is possible for
octagons and higher polygons. Since up to now the octagon remainder for
generic configurations is not available, our final result will be a substantial
prediction for a certain limiting behaviour of this unknown function. In this
context one should note that the self-crossing limit cannot be reached
within the special octagon configurations for which an analytical result
has been obtained in  \cite{duhr-8}. Furthermore, the anticipated effect
of conformal invariants on the prefactor of the self-crossing related 
divergence indicates, that the origin of this  term in the
wanted exact remainder function should be visible already without handling 
the subtleties of analytic continuation.

The logic of our paper will follow the lines of \cite{georgiou}. The bare 
(dimensionally regularised) Wilson loop is given by the BDS structure plus 
a remainder function $\mathcal R$. In a generic non-intersecting configuration the
remainder in the limit $\epsilon\rightarrow \nolinebreak 0$ remains finite, becomes independent
of the RG scale $\mu$ and depends on conformal invariants of the polygon only.
Then it constitutes a part of the renormalised Wilson loop for
non-intersecting configurations. Since new divergences appear in a configuration 
with self-crossing, we expect corresponding short distance singularities
in the limit of configurations with  self-crossing, both in the well-known 
contributions from  the BDS structure and in the unknown remainder function.
Our goal is to find the singularity for the remainder function $\mathcal R$.

To proceed in this direction we study the RG equation for the Wilson loop
in a self-crossing configuration. Then the remainder function has poles in
$\epsilon $. What remains after subtraction of these poles as contribution
to the renormalised self-crossing Wilson loop we call $\mathcal R_{\text{ren}}$.
\footnote{The index ``$\text{ren}$'' will be used  to mark the renormalised quantities in the self-crossing situation only.} Inserting the known  BDS structure one ends with an equation for
$\mathcal R_{\text{ren}}$, which fixes the dependence on powers of $\log\mu$.
Since in dimensional regularisation $\mu$ originates exclusively as a factor
$\mu ^{2\epsilon}$ in combination with the coupling constant $g^2$, one then can
conclude backward, which poles in $\epsilon$ the remainder $\mathcal R$ has in
the self-crossing situation. The final step will be based on the usual observation
that the leading singularities in dimensional regularisation and point splitting
regularisation coincide, if $\frac{1}{\epsilon}$ is identified with the logarithm 
of the distance.      
\section{RG equation for Wilson loops with self-crossing and cusps}
We are interested in Wilson loops for null polygons with $n\geq 8$ vertices
$x_1,\dots x_n$. For this purpose we first start with polygons $\mathcal C$, which 
are not of null type, i.e. $p_j^2\neq 0,\\p_j=x_{j+1}-x_j$, and discuss the
light-like limit afterwards. The Mandelstam variables are defined as $s_{jk}=(x_j-x_k)^2$. Let two vertices $x_{\hat k}$ and $x_{\hat l}$ coincide,
with more than two vertices between $x_{\hat k}$ and $x_{\hat l}$ on both parts of the polygonal contour ${\cal C}={\cal C}_{\hat k \hat l}\cdot {\cal C}_{\hat l\hat k}$, see fig.1. Then, with
$~~{\cal U}({\cal C})=\frac{1}{N}\mbox{tr}P\exp \left (ig\int _{{\cal C}}A_
{\mu}dx^{\mu}\right )$ in $SU(N)$ gauge theory,
\be
{\cal W}_1 ~=~\langle ~{\cal U}({\cal C})~\rangle ~~~~\mbox{and}~~~~
{\cal W}_2 ~=~\langle ~{\cal U}({\cal C}_{\hat k\hat l})~{\cal U}({\cal C}_{\hat l\hat k})~\rangle \label{loops}
\ee
mix under renormalisation \cite{pol},\cite{brandt,dorn,korch2} 
\be
\mathcal W_b~=~ Z_{bc}~ Z ~\mathcal W_c^\text{ ren}~.\label{WZW}
\ee
Here $Z$ is the product of the $Z$-factors for the cusps
at the vertices $x_l,~~l\neq \hat k,\hat l$ and the matrix $ Z_{bc}$ takes care
of the UV divergences at the crossing point $x_{\hat k}=x_{\hat l}$. From \eqref{WZW}
one gets in standard manner the RG equation (the $\beta$ function is zero
for $\mathcal N=4$ SYM)
\be
\mu\frac{\partial}{\partial\mu}~ \mathcal W_a^\text{ren}~=~-\Gamma_{ab}(g^2,\{\vartheta \}_{\text{cross}})~\mathcal W_b^\text{ren}~-\sum _{k
\neq \hat k,\hat l}\Gamma_{\text{cusp}}(g^2,\vartheta _{k,k-1})~\mathcal W_a^\text{ren}~.
\label{RG}
\ee
The angles $\vartheta _{k,l}$ are defined by
\footnote{We use the  $i0$-prescription as induced from that of the standard gluon propagator in position space. It has been argued, that for Wilson loops in correspondence
to scattering amplitudes the sign has to be reversed \cite{georgiou,brandhuber}.} 
 $~\cosh \vartheta _{k,l}=~
\frac{p_kp_l-i0}{\sqrt{(p_k^2-i0)(p_l^2-i0)}}~$ and  $~\{\vartheta \}_{\text{cross}}~$ stands for the six angles at the point of self-intersection.
\\*The anomalous dimension  matrix $\Gamma_{bc}$ is related to the matrix 
  $Z_{bc}$  via 
$\Gamma _{bc} = \mu \frac{\partial}{\partial \mu } (\log  Z)_{bc}$ and has been calculated in \cite{korch2} up to second order in QCD for a smooth intersection. We are interested in the case where we have two cusps at the intersection  point. Direct one-loop calculation leads to  \footnote{For a smooth intersection, i.e.
$p_{\hat k}=\lambda p_{\hat k -1},~~p_{\hat l}=\kappa p_{\hat l-1},~~\lambda ,\kappa >0$ one gets back the matrix found in \cite{korch2} (after adapting the normalisation
of $\mathcal W_2$ according to \eqref{loops}, see also comments on this in 
\cite{georgiou}).}
\bea
\Gamma_{11}&=& \frac{g^2 }{8\pi^2}\left ( \frac{N^2-1}{N}\big (f_{\hat l-1,l}+f_{\hat k-1,\hat k}-2\big ) -\frac{1}{N}\big (B_1~+~i \pi ~ h_{\hat k,\hat l}\big )\right )\nonumber\\
\Gamma_{22}&=& \frac{g^2 }{8\pi^2}\left (\frac{N^2-1}{N}\big (f_{\hat k,\hat l-1}+f_{\hat k-1,\hat l}-2\big )
-\frac 1 N \big (B_2~+~i \pi ~ h_{\hat k,\hat l}\big )\right )\nonumber\\
\Gamma_{12}&=& \frac{g^2 }{8\pi^2} N \big (B_1~+~i \pi h_{\hat k,\hat l}\big )\nonumber\\
\Gamma_{21}&=& \frac{g^2 }{8\pi^2}\frac 1 N \big (B_2~+~i \pi h_{\hat k,\hat l} \big )~,\label{Gamma}
\eea
 with the abbreviations 
\bea
f_{k,l}&:= &\vartheta_{kl}\coth \vartheta_{kl}~,~~~~~
h_{\hat k,\hat l}~:=~\coth \vartheta_{\hat k-1,\hat l-1}+\coth \vartheta_{\hat k,\hat l}
\nonumber \\
B_1&:=&f_{\hat k,\hat l-1}+f_{\hat k-1,\hat l}-f_{\hat k,\hat l}-f_{\hat k-1,\hat l-1}\nonumber\\
B_2&:=&f_{\hat k -1,\hat k}+f_{\hat l-1,\hat l}-f_{\hat k,\hat l}-f_{\hat k-1,\hat l-1}~.
\label{abbrev}
\eea

\begin{figure}
 \centering
 \includegraphics[width=4cm,bb=168 218 427 721]{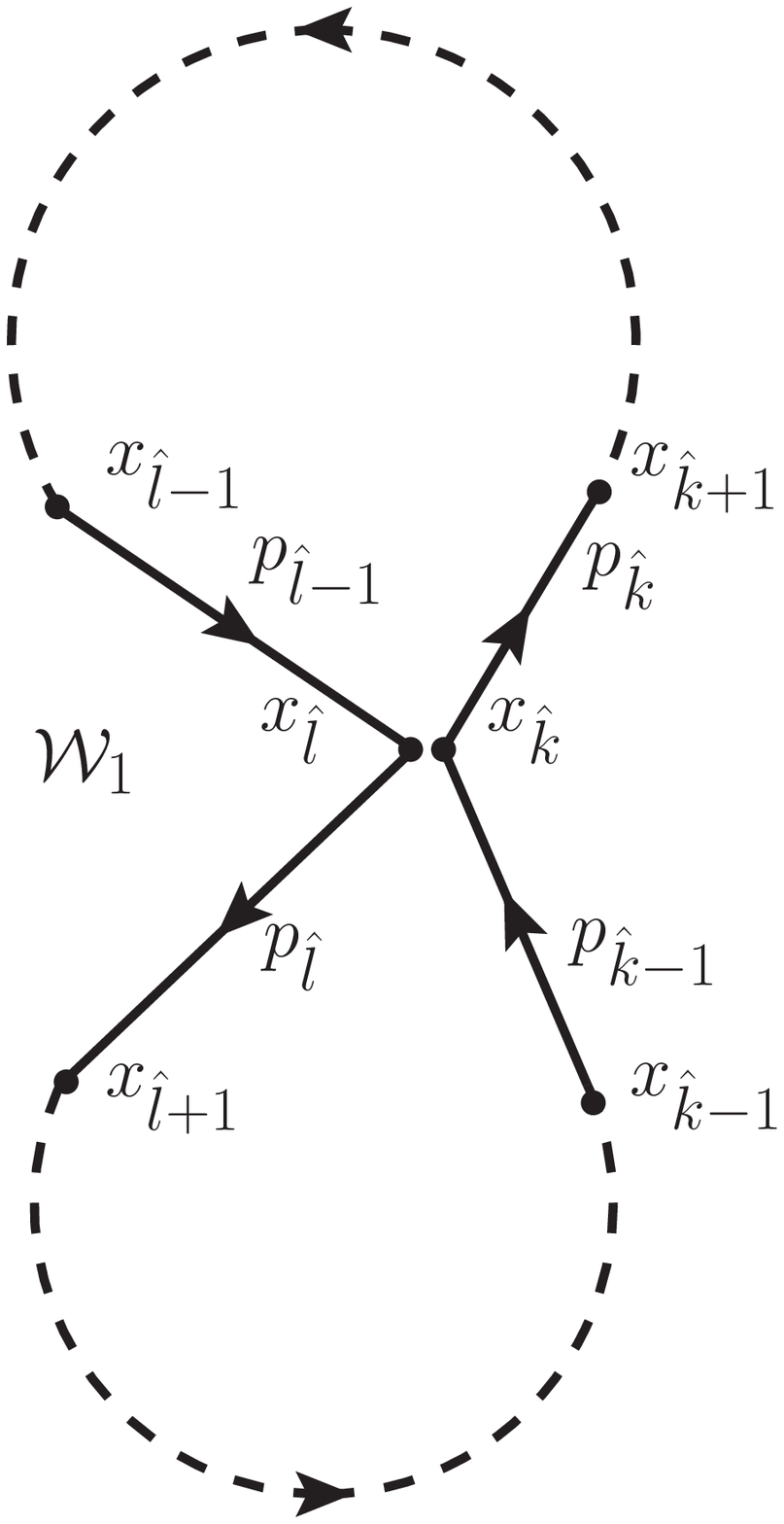} \hspace{30pt}
 \includegraphics[width=4cm,bb=168 218 427 721]{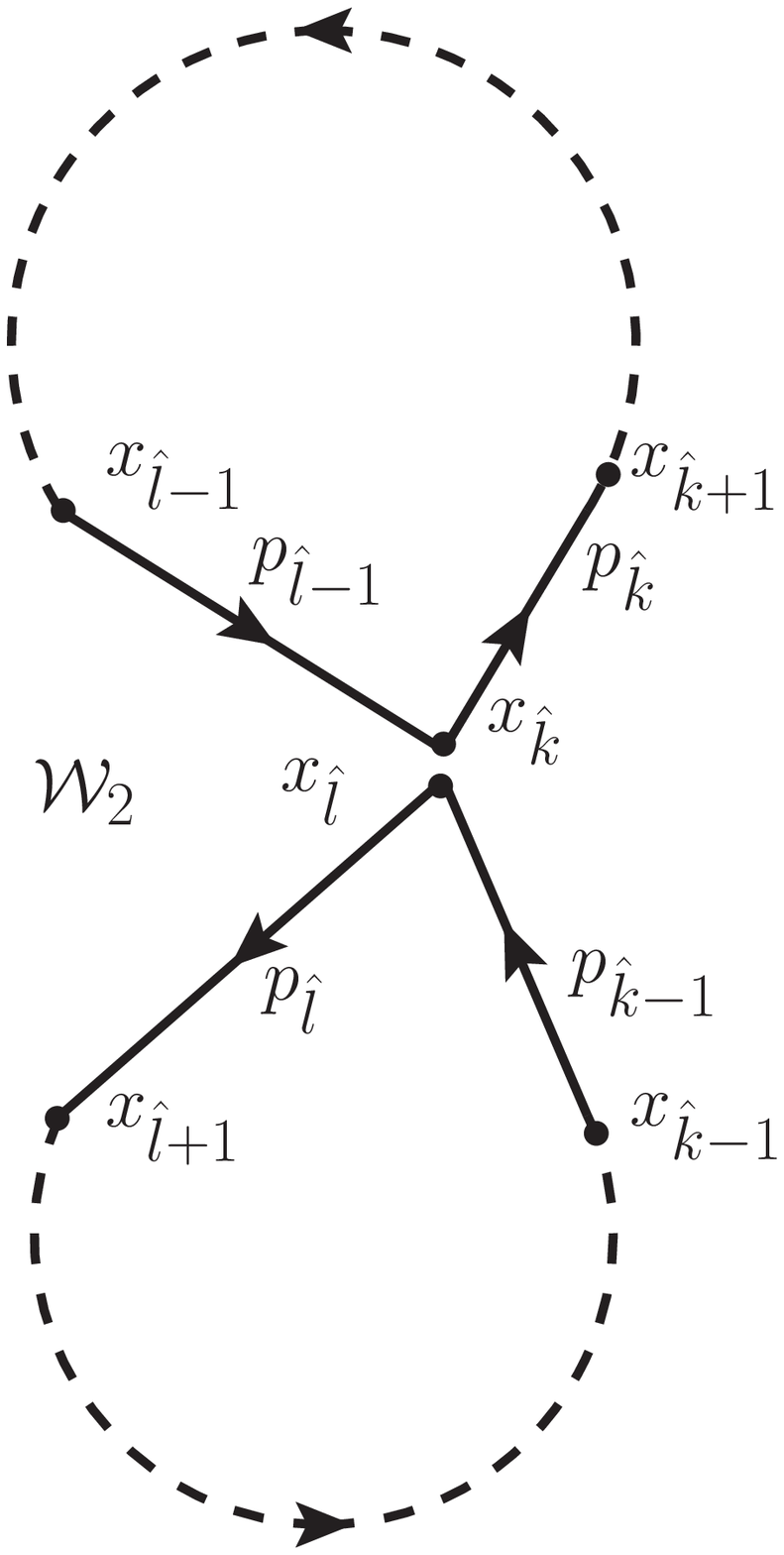}
\caption{The crossing situation for $x_{\hat k}=x_{\hat l}$. Both Wilson loops mix under renormalisation. The dashed lines correspond to some further arbitrary light-like polygonal segments of the Wilson loop.}
 \label{fig1}
\end{figure}

We now turn to the light-like limit $p_k^2\rightarrow 0, \forall k$. 
Then all angles $\vartheta _{k,l}$ diverge like
\be
\vartheta _{k,l}~=~\log\frac{2p_kp_l-i0}{\sqrt{(p_k^2-i0)(p_l^2-i0)}}~
\label{approxangle}~,
\ee
and their hyperbolic cotangent can be replaced by $1$. 
This implies that $B_1$ and $B_2$ become logarithms of cross ratios
\be
B_1~+~2\pi i~=~\log\frac{s_{\hat k+1, \hat l-1}~s_{\hat k-1, \hat l+1}}{s_{\hat k+1, \hat l+1}~ s_{\hat k-1, \hat l-1 }} ~,~~~~B_2~+~2\pi i~=~\log\frac{s_{\hat k-1, \hat k+1}~s_{\hat l-1, \hat l+1}}{s_{\hat k+1, \hat l+1}~s_{\hat k-1, \hat l-1}}~.\label{Capprox}
\ee
 
In general there are two independent cross ratios associated with four points. In four dimensions the position of $x_{\hat k} = x_{\hat l}$ is fixed by the light-likeness condition for the  four neighbouring  edges and the  neighbouring  points $\{x_{\hat k+1}, ~ x_{\hat l-1},~ x_{\hat l+1},~ x_{\hat k-1}\}$ are not restricted. Thus there are two independent cross ratios describing the crossing situation that we encounter in the matrix
$\Gamma_{bc}$.

Finally,   performing   the    't Hooft limit $N \rightarrow \infty$, with $a:= \frac{g^2 N}{8 \pi^2}$ kept fixed, we arrive at  
\begin{equation}
\Gamma_{bc}= a \begin{pmatrix} \vartheta_{\hat l-1,\hat l}+\vartheta_{\hat k-1,\hat k} -2 & B_1 +2\pi i \\ 0 & \vartheta_{\hat k,\hat l-1}+\vartheta_{\hat k-1,\hat l}-2 \end{pmatrix}~+\mathcal O(a^2)~.\label{Gammaplanar}
\end{equation}
Note that due to the colour structure $Z_{21}$ and $\Gamma_{21}$ are zero in all  orders of perturbation theory.

In the light-like limit  $\Gamma _{11},~\Gamma_{22}$ and $\Gamma_{\text{ cusp}}(g^2,\vartheta_{k,k-1})$ become divergent and make the RG equation \eqref{RG} ill defined.
According to \cite{korchrad,korch1} the anomalous dimension  $\Gamma_{\text {cusp}}(g^2,\vartheta)$ for large $\vartheta$ has an all order asymptotic behaviour
$\Gamma_{\text{cusp}}(g^2,\vartheta)=\vartheta ~\Gamma _{\text{cusp}}(a)  +\mathcal O(1)$. Based on this observation in \cite{korch1}, by suitable differentiation with 
respect to Mandelstam variables and backward integration, a modified RG equation
has been derived for Wilson loops for non-intersecting null polygons. The resulting 
equation can be described by the following recipe: keep the structure of the RG equation and replace every vanishing $p_k^2$ by $-\frac{1}{\mu ^2}$, where $\mu$ is the
RG scale. In the process of backward integration a new integration constant appears. It  depends  on $g^2$ only.
The equation has been checked explicitly on two loop level \cite{korchmarch}.
Following \cite{georgiou} we assume the same recipe to work also in the case of 
Wilson loops for self-crossing null polygons. An analogous structure has been
obtained in the study of infrared divergences of scattering amplitudes \cite{soft}. 

Our basic RG equation for $\mathcal W_1^\text{ren}$, obtained with the just described
procedure from \eqref{RG},\eqref{Gammaplanar},\eqref{approxangle}  is then
\be
\mu \frac{\partial}{\partial \mu} \log \mathcal W_1^\text{ren} = -\Gamma_{12}~ \frac{\mathcal W_2^\text{ren}}{\mathcal W_1^\text{ren}} -\Big ( \Gamma_{11}+ \bar \Gamma (a) +\frac{\Gamma_{\text{cusp}}(a)}{2}\sum_{k \neq \hat k,\hat l}  \log(-\mu^2 s_{k-1,k+1}+ i0)  \Big )\label{basic}~.
\ee
$\Gamma_{\text{cusp}}$ and the new object $\bar \Gamma $, arising in the approach of \cite{korch1} as integration constant, depend on the coupling $a$ only. 
$\Gamma_{12}=a(B_1+2\pi i)+\mathcal O (a^2)$. For convenience
we understand the $p_k$ independent part of $\Gamma_{11}$ to be included in $\bar\Gamma$ and will use
\be
\Gamma_{11}~=~ a~\big (\log(-\mu^2 s_{\hat l-1,\hat l+1} + i0)+\log(-\mu^2 s_{\hat k-1, \hat k+1} + i0)\big )+\mathcal O (a^2)~.
\label{g11}
\ee 
The crucial property of \eqref{basic} is, that since $\Gamma _{12}$ starts  at  order
$a$, to balance the order $a^2$ of  $\log \mathcal W_1^\text{ren}$ , only one loop information on  $ \mathcal W_1^\text{ren},\mathcal W_2^\text{ren}$  is needed on the right hand side.
 \section{BDS structure and RG equation for the \\remainder} 
Taking into account the recursive BDS structure \cite{bds,anastasiou}, corrected
by the remainder function $\mathcal R_n$, the  generic  
$n$-sided null polygon Wilson loop is given by
\be
\log \mathcal W~=~\sum _{l=1}^{\infty}~a^l\big ( f^{(l)}(\epsilon)~w(l\epsilon)~+~C_n^{(l)}\big )~+~
\mathcal R_n~+~\cal O(\epsilon)~.\label{bds}
\ee
Here the $C^{(l)}$ are numbers, $f^{(l)}(\epsilon)=f^{(l)}_0+\epsilon f^{(l)}_1+\epsilon^2f^{(l)}_2$, and $w_n(\epsilon)$ is the one loop contribution
\be
w_n(\epsilon)~=~-\frac{1}{2}~\sum_{k=1}^n\frac{1}{\epsilon ^2}~(-\mu ^2 s_{k-1,k+1}
  +   i0)^{\epsilon} ~+~F_n(\mu ^2,\epsilon, s)~.\label{1loop}
\ee
For a generic null polygon configuration $F_n$ and $\mathcal R_n(\mu^2,\epsilon,s)=a^2\mathcal R^{(2)}_n(\mu^2,\epsilon,s)+\dots$ stay
finite and become independent of $\mu ^2$ in the limit $\epsilon\rightarrow 0$.\\
Relating $f^{(l)}$ via
\be
f^{(l)}_0~=~\frac{\Gamma^{(l)}_{\text{cusp}}}{2}~,~~~~f^{(l)}_1=\frac{l~\Gamma^{(l)}}{2}\label{f-gamma}
\ee
to the cusp anomalous dimension and the collinear anomalous dimension, as well as taking into account $f^{(1)}(\epsilon)=1$  (i.e. $\Gamma_{\text{cusp}}^{(1)}=2$),    $C^{(1)}=0$ and $\Gamma^{(1)}=0$
we get up to two loops \cite{georgiou} 
\bea
\log \mathcal W=-\frac{1}{4}\sum_{l=1,2}a^l\left (\frac{\Gamma^{(l)}_{\text{cusp}}}{(l\epsilon)^2}+\frac{\Gamma ^{(l)}}{l\epsilon}\right )\sum_k  (-\mu^2s_{k-1,k+1})^{l\epsilon}  ~+a~F_n(\mu^2,\epsilon,s)-\frac{a^2n}{8}f^{(2)}_2\nonumber\\
 ~~~+a^2\Big (\frac{\Gamma^{(2)}_{\text{cusp}}}{2}F_n(\mu^2,2\epsilon,s)
 +\epsilon\Gamma^{(2)} F_n(\mu ^2,2\epsilon,s)+C^{(2)}+\mathcal R^{(2)}_n(\mu^2,\epsilon,s)\Big )+{\cal O}(\epsilon).\label{2loopbds}
\eea
The term $\epsilon \Gamma^{(2)}F_n$ has been kept, since in the crossing configuration
under discussion $F_n$ develops a pole in $\epsilon$. As a consequence, now the ${\cal O}(\epsilon , a^3)$ estimate holds not only in the generic, but also in the limit of a configuration with crossing. 
\begin{figure}
 \centering
 \includegraphics[width=4cm,bb=185 515 421 721]{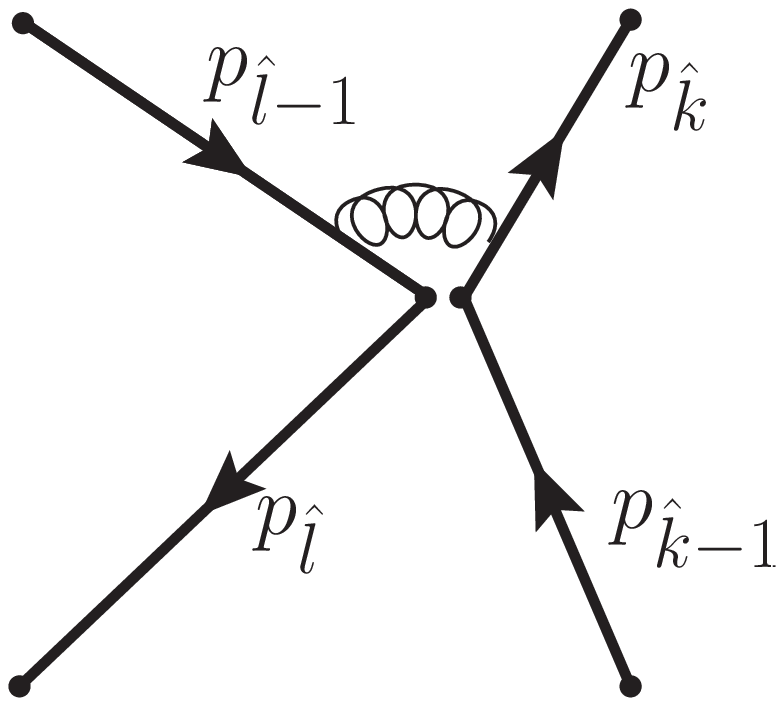} \hspace{30pt} \includegraphics[width=4cm,bb=185 515 421 721]{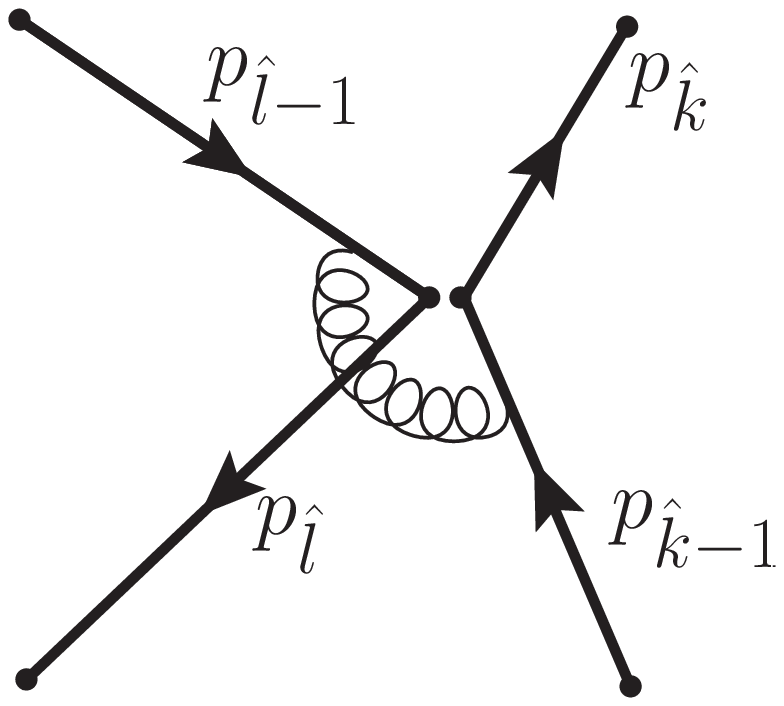}
\caption{ These are two of four diagrams that are responsible for the new divergences in $F_n$. }
 \label{fig:fin}
\end{figure}

There are three sources for pole terms. The poles
of the first term on the r.h.s. are present already in a generic configuration.
After expanding the terms  $(-\mu ^2s_{k-1,k+1})^{l\epsilon}$   one gets $\log ^2$ and
$\log$ terms in momenta as contributions to   $\log \mathcal  W^{\text{ren.}}$. 
The remainder function
becomes divergent in the crossing configuration, let us call $\mathcal R^{(2)\text{ren}}_n(\mu^2,s)$ 
what remains after subtraction of the poles in $\epsilon$. The last source for pole terms
is the one loop function $F_n$.
The poles of the one loop function $F_n$ in the crossing configuration arise 
from the diagrams  in fig. \ref{fig:fin}. 
Note that  these  diagrams  are  finite for a generic configuration and that the generic
poles of the one loop Wilson loop are taken into account by the first term of the
r.h.s. of \eqref{2loopbds} already. We find 
 \bea\label{f8} 
 F_n(\mu ^2,\epsilon,s)&=&\frac{1}{2\epsilon}~\log\frac{s_{\hat k-1, \hat l-1}~s_{\hat k+1, \hat l+1}}{s_{\hat k+1, \hat l-1}~s_{\hat k-1,\hat l+1}}\\
&+&\frac{1}{4}\big ( L^2_{\hat k - 1, \hat l - 1} +L^2_{\hat k + 1, \hat l + 1} -L^2_{\hat k + 1, \hat l - 1} - L^2_{\hat k - 1, \hat l + 1}\big ) ~+~\tilde F_n (s)~+~{\cal O}(\epsilon)~, \nonumber
\eea 
where $\tilde F_n (s)$ is now independent of $\mu^2$. In order to improve the 
readability of our formul\ae, we introduced the following abbreviation 
\footnote{For notational convenience we drop the $i0$ terms later on.
$s_{kl}$ stands for $s_{kl}-i0$.}
\begin{equation}
L_{jk} := \log (-\mu^2 s_{jk}+i0)~.
\end{equation}

We now extract from \eqref{2loopbds} and \eqref{f8} all the ingredients for the RG 
equation \eqref{basic} and start with
the quotient $\frac{\mathcal W_2^{\text{ren}}}{\mathcal W_1^{\text{ren}}}$, which will be needed
in one loop approximation only. In this order $\log \mathcal W_1^{\text{ren}}$
is given by minimal subtraction of corresponding poles in $\epsilon$ in \eqref{2loopbds} taking into account \eqref{f8}  
\begin{align}
 \label{W1-1}
\log \mathcal  	W_1^{\text{ren}}~=~-~&\frac{a}{4} \Big  (L^2_{\hat k+1,\hat l-1}+L^2_
{\hat k-1,\hat l+1}-L^2_{\hat k-1,\hat l-1} -L^2_{\hat k+1,\hat l+1}+ \sum_{k=1}^n 
L^2_{k-1,k+1} \Big ) \nonumber \\
+~& a~\tilde F_n(s) ~+~\mathcal O(a^2)~.
\end{align}  
In the planar limit under discussion, $\mathcal W_2$ for the self-crossing $n$-gon
factorises in the product of two Wilson loops for the two parts, the $n_+$-gon $\mathcal C_{\hat k\hat l}$ and the $n_-$-gon $\mathcal C_{\hat l\hat k}$ ($n_++n_-=n$). For these two factors \eqref{f8} is irrelevant and we get from \eqref{2loopbds}
\bea\label{W2-1}
\log \mathcal W_{2}^{\text{ren}}& ~=~ &-~\frac{a}{4}\big ( L^2_{\hat k,\hat k+2}  + \dots +L^2_{\hat k+1, \hat l-1}\big ) ~+~a F_{n_+}   \\
&&-~\frac{a}{4}\big (L^2_{\hat l, \hat l+2}  + \dots +L^2_{\hat k-1,\hat l+1}\big ) ~+~a F_{n_-} ~+~ \mathcal O(a^2) ~.\nonumber  
\eea
Together with \eqref{W1-1} the last equation implies

\begin{align}\label{2div1}
\frac{\mathcal W_2^{\text{ren}}}{\mathcal W_1^{\text{ren}}} = & ~ 1~+~ \frac{a}{4}\Big ( L^2_{\hat l-1 , \hat l+1} +L^2_{\hat k-1 , \hat k+1} -L^2_{\hat k-1, \hat l-1 }- L^2_{\hat k+1 ,\hat l+1} \Big) \nonumber\\
+&~a~(F_{n_+}+F_{n_-}-\tilde F_n)~+~\mathcal O(a^2)~.
\end{align}
$F_{n_+},~F_{n_-}$ and $\tilde F_n$ are independent of $\mu ^2$ for $\epsilon =0$.
For the l.h.s. of \eqref{basic} we need $\log \mathcal W_{1}^{\text{ren}}$ at order 
$a^2$. If there  was  no mixing with $\mathcal W_2$ we would get   the $\mathcal O(a^2)$  contribution, similar to the  lowest 
order, by minimal subtraction of $\epsilon $-poles in  \eqref{2loopbds},\eqref{f8}.
To take care of the mixing effect, let us denote ($b=1,2$)
\be
\mathcal V_b:=\log W_b=\sum _j a^j ~\mathcal V_b^{(j)}
\label{v}
\ee
and use similar power expansions for $\mathcal V _b^{\text{ren}}$ and $\mathcal Z _{bc}:=Z_{bc}~Z$. Then \eqref{WZW} implies
\bea
\mathcal V^{\text{ren}(1)}_1 &= &\mathcal V^{(1)}_1 -
\mathcal Z^{(1)}_{11}-\mathcal Z^{(1)}_{12}\nonumber\\
\mathcal V^{\text{ren}(2)}_1& =& \mathcal V^{(2)}_{1}  + 
  \mathcal Z_{12}^{(1)} \left( \mathcal V^{ \text{ren}(1)}_{1} -\mathcal V^{\text{ren}(1)}_{2} \right)~\nonumber \\
&-&\mathcal Z_{11}^{(2)} -\mathcal Z_{12}^{(2)}+\mathcal Z_{11}^{(1)}\mathcal Z_{12}^{(1)} +\frac{1}{2} \left( \Big (\mathcal Z^{(1)}_{11} \Big )^2 + \Big ( \mathcal Z^{(1)}_{12} \Big )^2\right )~.\label{vz}
\eea
Therefore, $\mathcal V^{\text{ren}(2)}_1$ is given by the minimally subtracted 
first line of the r.h.s. of \eqref{vz} and consequently $\log \mathcal W_1^{\text{ren}}$ by the minimally subtracted r.h.s. of \eqref{2loopbds}, with \eqref{f8} in mind, {\it plus} 
\be
a^2~\lim_{\epsilon\rightarrow 0}\mathcal Z_{12}^{(1)} \left( \mathcal V^{ \text{ren}(1)}_{1} -\mathcal V^{\text{ren}(1)}_{2} \right)=-\frac{a^2}{2}~\Gamma_{12}^{(1)}~\frac{\partial}{\partial\epsilon}\left (
\mathcal V^{ \text{ren}(1)}_{1} - \mathcal V^{\text{ren}(1)}_{2}  \right)\vert _{\epsilon =0}~.\label{addit}
\ee 
Use has been made of $ \mathcal Z_{12}^{(1)}= Z_{12}^{(1)}=-\frac{1}{2\epsilon}~\Gamma_{12}^{(1)}$. \footnote{Note that the contribution from \eqref{addit} has been omitted in 
\cite{georgiou}. However, taking it properly into account would 
modify that result  at the end by a factor 2 only.} Similar to the derivation of 
\eqref{2div1} we get
\begin{align}\label{zusatz}
a^2 \lim_{\epsilon\rightarrow 0}\mathcal Z_{12}^{(1)} \Big ( \mathcal V^{ \text{ren}(1)}_{1} -~\mathcal V^{\text{ren}(1)}_{2} \Big )&= -\frac{a^2\Gamma_{12}^{(1)}}{24} \Big ( L^3_{\hat k-1, \hat l-1} +L^3_{\hat k+1, \hat l+1} -L^3_{\hat k-1, \hat k+1} -L^3_{\hat l-1, \hat l+1} \Big ) \nonumber \\
& +~a^2\cdot(\text{terms}\propto \log \mu^2) ~.
\end{align}
Now with \eqref{2loopbds},\eqref{f8} and \eqref{zusatz} we arrive at
\begin{align}\label{renW1}
\log \mathcal W_1^{\text{ren}} &= a\cdot(\dots) ~-~\frac{a^2\Gamma_{12}^{(1)}}{24} \Big (L^3_{\hat k-1, \hat l-1} + L^3_{\hat k+1, \hat l+1}- L^3_{\hat k-1, \hat k+1} - L^3_{\hat l-1, \hat l+1}\Big )  \nonumber\\
&-\frac{a^2\Gamma^{(2)}_{\text{cusp}}}{8}~ \Big  ( L^2_{\hat k+1, \hat l-1} + L^2_{\hat k-1, \hat l+1} -L^2_{ \hat k-1,\hat l-1}- L^2_{\hat  k+1,\hat l+1}+
\sum_{k=1}^n L^2_{k-1,k+1} \Big ) \nonumber\\
&+a^2\cdot (\text{terms}\propto \log\mu ^2)~+~a^2~\mathcal R_n^{(2)\text{ren}}(\mu^2,s)~+~ \mathcal O(a^3)~.
\end{align}
Inserting this together with \eqref{2div1} into \eqref{basic} and balance the order $a^2$ terms we get a RG equation for $\mathcal R_n^{(2)\text{ren}}$ (by this we denote  the minimally subtracted part of $\mathcal R_n^{(2)}$ here). 

\begin{align}\label{RGR}
\mu \frac{\partial}{\partial \mu}\mathcal R_n^{(2) \text{ren}}&=\frac{\Gamma^{(2)}_{\text{cusp}}}{2}~ \Big ( L_{\hat k-1, \hat k+1} + L_{\hat l-1, \hat l+1} + L_{\hat k+1,\hat l-1} + L_{\hat k-1, \hat l+1} - L_{\hat k-1, \hat l-1} - L_{\hat k+1,\hat l+1} \Big )\nonumber\\
&+\frac{\Gamma_{12}^{(1)}}{2} \Big ( L^2_{\hat k-1, \hat l-1} + L^2_{\hat k+1, \hat l+1} -L^2_{\hat k-1, \hat k+1} -L^2_{\hat l-1, \hat l+1} \Big )~-~ \Gamma^{(2)}_{11}~+~\dots ~,
\end{align}
where the dots stand for terms independent of $\mu ^2$. They include $-\Gamma ^{(2)}_{12} $, which as $-\Gamma ^{(1)}_{12} $ should be independent of $\mu ^2$.
The only interesting unknown entry on the r.h.s. is $\Gamma^{(2)}_{11}$.  We expect the situation to be  similar   to the cusp anomalous dimension, where in the light-like
limit one   can factor off  a linear dependence on  $\log (-2\mu^2 p_{k-1}p_k)$.  Assuming
  such a    behaviour also for the crossing matrix entries,	
 we get   $\Gamma_{11}^{(2)} = \gamma_{11}^{(2)}\big ( L_{\hat k-1,\hat k+1} + L_{\hat l-1,\hat l+1} \big ) $
with a number $\gamma_{11}^{(2)}$ that has to be determined in a two-loop 
calculation.   Then integration of \eqref{RGR} yields
\begin{align}\label{rem}
\mathcal R_n^{(2)\text{ren}}&=\frac{\Gamma^{(2)}_{\text{cusp}}}{8}~ \Big  (    L^2_{\hat k-1, \hat k+1} + L^2_{\hat l-1, \hat l+1} + L^2_{\hat k+1,\hat l-1} + L^2_{\hat k-1, \hat l+1} - L^2_{\hat k-1, \hat l-1} - L^2_{\hat k+1,\hat l+1}   \Big )\nonumber\\
&+\frac{\Gamma_{12}^{(1)}}{12} \Big (  L^3_{\hat k-1, \hat l-1} + L^3_{\hat k+1, \hat l+1} -L^3_{\hat k-1, \hat k+1} -L^3_{\hat l-1, \hat l+1}  \Big ) \\
&- \frac{\gamma^{(2)}_{11}}{4}\Big (L^2_{\hat k-1,\hat k+1} + L^2_{\hat l-1,\hat l+1}\Big ) ~+~\mathcal O(\log \mu ^2)~. \nonumber
\end{align}
This is the  two loop   remainder function   renormalised to accommodate
the extra divergences due to the self-crossing. Since $\mu ^2$ in the 
dimensionally regularised $\mathcal R_n^{(2)\text{ren}}$ originates from the 
expansion of $g^4\mu ^{4\epsilon}$, one 
can backward reconstruct the unrenormalised remainder function 
\begin{align}\label{Rbare}
\mathcal R_n^{(2)}(\mu ^2 ,\epsilon,s)&=\frac{\Gamma^{(2)}_{\text{cusp}}}{16\epsilon ^2}~ \Big  ((-\mu^2 s_{\hat k-1, \hat k+1})^{2\epsilon}+(-\mu^2 s_{\hat l-1, \hat l+1})^{2\epsilon} +(-\mu^2 s_{\hat k+1, \hat l-1})^{2\epsilon}\nonumber\\
&~~~~~~~~~~~~+(-\mu^2 s_{\hat k-1, \hat l+1})^{2\epsilon}-(-\mu^2 s_{\hat k-1,\hat l-1})^{2\epsilon}-(-\mu^2 s_{\hat k+1, \hat l+1})^{2\epsilon} \Big )\nonumber\\
&+\frac{\Gamma_{12}^{(1)}}{16\epsilon ^3} \Big ( (- \mu^2 s_{\hat k-1, \hat l-1})^{2\epsilon}+
 (- \mu^2 s_{\hat k+1, \hat l+1})^{2\epsilon}\nonumber\\
&~~~~~~~~~~~~-(-\mu^2 s_{\hat k-1, \hat k+1})^{2\epsilon}- (- \mu^2 s_{\hat l-1, \hat l+1})^{2\epsilon} \Big )\nonumber\\
&~-~ \frac{\gamma^{(2)}_{11}}{8\epsilon ^2}\Big ((-\mu^2 s_{\hat k-1, \hat k+1} )^{2\epsilon}+ (-\mu^2s_{\hat l-1, \hat l+1} )^{2\epsilon} \Big )~+~\mathcal O(\frac{1}{\epsilon})~.
\end{align}
Expanding the exponents and  inserting $\Gamma_{12}^{(1)}$ from \eqref{Capprox},\eqref{Gammaplanar} we finally get
\be\label{final}
\mathcal R_n^{(2)} = \frac{1}{8\epsilon^2} \Big  ( \log \frac{s_{\hat k-1,l+1}s_{\hat k+1,\hat l-1}}{s_{\hat k-1,\hat l-1}s_{\hat k+1,\hat l+1}}
~ 
\log  \frac{ s_{\hat k-1,\hat l-1}  s_{\hat k+1,\hat l+1}}{s_{\hat l-1,\hat l+1}
s_{\hat k+1,\hat k-1}}  - 2\gamma_{11}^{(2)} + \Gamma_{\text{cusp}}^{(2)}\Big )+
\mathcal O(\frac{1}{\epsilon})~.
\ee
If instead of dimensional regularisation one uses a point splitting
regularisation \\$x_{\hat k}=x_{\hat l}+\delta \cdot v$ ($v$ some unit vector)
the leading divergences coincide, if one identifies\\ $\frac{1}{\epsilon^2}$
with   $\log^2(1/\delta^2)$.

Therefore, the two loop remainder function for a null n-gon, while  being finite
in generic configurations, develops a $\log ^2$ divergence in the distance, if two
vertices approach each other. The prefactor of this divergence depends on
cross-ratios formed out of the four neighbour vertices and is given by 1/8 times
the expression in brackets in \eqref{final}. 

For notational shortness, above we have been sloppy with indicating all the 
arguments
on which the remainder in different formul\ae~ depends. We end this section by 
summarising the complete pattern:
\bea
\mathcal R^{(2)}_n(\mu ^2,\epsilon,s_{kl})&=&\mathcal R ^{(2)}_n(u_{kl})~+
~\mathcal O(\epsilon) \nonumber\\
\mathcal R^{(2)}_n(\mu ^2,\epsilon,\{s_{kl}\})&=&\frac {1}{\epsilon ^2}~H_n(\{u_{kl}\})~
+~\frac {1}{\epsilon}(\dots )~+~\mathcal R_n^{(2)\text{ren}}(\mu ^2,\{s_{kl}\})~+
~\mathcal O(\epsilon)\nonumber\\
\mathcal R ^{(2)}_n(u_{kl})&=& \log ^2 (\delta^2)  ~G_n(\{ u_{kl}\})~+
~\mathcal O(\log\delta ^2  )~.
\eea
$\{s_{kl}\}$ and $\{u_{kl}\}$ denote the set of Mandelstam variables and 
cross-ratios in the self-crossing limit. Finally we have used $H_n=G_n$. 

\section{Octagon }
As an example, we now specialise to the octagon and chose $\hat k=1$ and $\hat l=5$.
The configuration   of an octagon (in every dimension) can be described using at most $12$ conformally invariant cross ratios. In four dimension there are only $9$ independent cross ratios due to Gram constraints. So far it has not been possible to disentangle these relations for four dimensions. So we use the usual choice for the $12$ conformal cross ratios 
\begin{equation}
u_{ij} = \frac {(x_i-x_{j+1})^2(x_{i+1}-x_j)^2}{(x_i-x_j)^2(x_{i+1}-x_{j+1})^2}~.
\end{equation}
Let us    look at these cross ratios in the limit $x_1 = x_5 + \delta ~  v,   ~~\delta \rightarrow 0$,   when the loop 
  becomes  self-intersecting.   We want to express a divergence in $\delta$ as a divergence in terms of conformal invariants. This relation will also contain Mandelstam variables (which are not conformally invariant) because distances are also not conformally invariant.

In the aforementioned limit we encounter three classes of cross ratios. The ratios $
u_{26},u_{27},u_{36},  u_{37}   $ are not affected by this limit and remain untouched. Four cross ratios $u_{14},u_{15},u_{48},u_{58}$ remain finite (in the general case) but depend on the direction $v$.   For example one finds
\be
u_{14}~=~\frac{v^2(x_2-x_4)^2}{4vp_4~~vp_1}~.
\ee
 The last class diverges as we approach the crossing situation $u_{16},u_{25},u_{38},u_{47}$,   e.g.
\be
u_{16}~=~-\frac{1}{\delta}~\frac{(x_2-x_6)^2(x_1-x_7)^2}{2 vp_5~~(x_2-x_7)^2}~.
\ee 
We can eliminate the dependence on the direction of $v$ by considering combinations of various $u_{kl}$ and   
find the relation   
\begin{equation}\label{delta}
4 \log \delta =  -\log \big (u_{47}u_{38}u_{25}u_{16}\big )+\log \Big ( \frac {s_{48}s_{57}s_{13}s_{26}s_{35}s_{17}}{s_{47}s_{38}s_{36}s_{27}~ v^4} \Big ) -\log (u_{15}u_{48})
\end{equation}
for the crossing limit.
The first term on the r.h.s. of \eqref{delta} is conformally
invariant and becomes divergent in the limit. The other two terms stay finite
and balance the conformal non-invariance of the l.h.s.. 
Finally with the abbreviation $u:=u_{47}u_{38}u_{25}u_{16}$ we get from \eqref{final}
and the discussion at the end of the previous section
\be
\mathcal R_8^{(2)} =\frac{1}{32} \log^2 u ~\Big ( 
\log \frac{s_{86}s_{24}}{s_{84}s_{26}}
~ \log  \frac{s_{48}s_{26}}{s_{46}s_{28}}  - 2\gamma_{11}^{(2)} + \Gamma_{\text{cusp}}^{(2)}
\Big )~ +~ \mathcal O(\log u)\label{end}
\ee
for $x_1\rightarrow x_5$. This is valid as long as the vector $v$ defining the direction of the approach is not light-like and  has a nonzero scalar product 
with $p_1,p_4,p_5,p_8$. 

\section{Conclusions}
With RG techniques we have calculated the leading divergence of the two 
loop remainder function in the limit of two approaching vertices of the null
polygon. We found a behaviour $\propto 
\frac{1}{2}\log ^2\delta $, where $\delta$
measures the vanishing distance between the approaching vertices.
The prefactor of this divergence is given by the product of two logarithms
of cross-ratios parametrising the conformal geometry of the self-crossing
plus some pure number. Only the determination of this number requires
two loop calculations, all other ingredients are fixed by the well-known
one loop structure of the matrix of anomalous dimensions.

The prefactor itself becomes logarithmically divergent if the self-crossing
configuration degenerates to the crossing of two smooth pieces of the
Wilson loop contour. In the octagon case, for example, such a situation
would arise if $p_4$ and $p_5$ as well as $p_8$ and $p_1$ become collinear 
(i.e. $s_{46},~s_{28}\rightarrow 0$). This reflects the $\log ^3$ divergence 
found in \cite{georgiou} for self-intersections at interior points of the 
edges of the polygon. 

Our result could be checked independently by direct study of the
corresponding limit in the Feynman diagrams responsible for the extra
divergences in the self-crossing configuration. But even when the full two
loop remainder will be available, the RG technique again can be used
to get information on such special limits one order higher.\\[10mm]
{\it Note added:}\\
In the original version of this paper we had used another translation factor 
between dimensional and point splitting regularisation. This led in \eqref{end}
to a factor 1/128 instead of 1/32. Strong arguments for the translation
rule used now are given in our recent paper \cite{dw}.  
\\[10mm]
\noindent
{\bf Acknowledgement}\\[2mm]
We  thank  George Georgiou, Johannes Henn and Chrysostomos Kalousios for useful
discussions. This work was supported by DFG via GK 1504 and SFB 647.

\end{document}